\documentclass[twocolumn]{aastex701}
\usepackage[T1]{fontenc}
\usepackage{hyperref}
\usepackage{amsmath}
\usepackage{amssymb}
\usepackage{amsfonts}
\usepackage[dvipsnames]{xcolor}

\usepackage{hyperref}
	\hypersetup{
    unicode=true,          
    pdftoolbar=true,        
    pdfmenubar=true,        
    pdffitwindow=false,     
    pdfstartview={FitH},    
    pdftitle={My title},    
    pdfauthor={Author},     
    pdfsubject={Subject},   
    pdfcreator={Creator},   
    pdfproducer={Producer}, 
    pdfkeywords={keyword1, key2, key3}, 
    pdfnewwindow=true,      
    colorlinks=true,        
    linkcolor=RoyalBlue,     
    citecolor=MidnightBlue,     
    filecolor=MidnightBlue,     
    urlcolor=MidnightBlue       
}

\newcommand{\usmg}{USMg\textsc{ii}}
  
\newcommand{\lya}{Ly$\alpha$ }

\newcommand{\zabs}{$z_{\rm abs}$}
\newcommand{\kms}{$km s^{-1}$}

\newcommand{\HI}{\mbox{H\,{\sc i}}}

\newcommand{\OII}{\mbox{[O\,{\sc ii}]}}
\newcommand{\OIII}{\mbox{[O\,{\sc iii}]}}

\newcommand{\MgII}{\mbox{Mg\,{\sc ii}}}

\newcommand{\FeII}{\mbox{Fe\,{\sc ii}}}

\newcommand{\CaII}{\mbox{Ca\,{\sc ii}}}

\newcommand{\gotoqs}{GOTOQs}

\begin{document}

\title{Cosmic Chance Superpositions: Largest Catalog of Quasar-Galaxy Pairs at a Projected Separation of $D \lesssim 20$ kpc}

\author[orcid=0009-0004-4263-077X]{Labanya K. Guha}
\affiliation{Indian Institute of Astrophysics (IIA), Koramangala II, Bangalore 560034, India}
\email{labanya.guha@iiap.res.in}

\author{Raghunathan Srianand}
\affiliation{IUCAA, Postbag 4, Ganeshkhind, Pune 411007, Maharashtra, India}
\email{anand@iucaa.in}

\author{Rajeshwari Dutta}
\affiliation{IUCAA, Postbag 4, Ganeshkhind, Pune 411007, Maharashtra, India}
\email{rajeshwari.dutta@iucaa.in}

\correspondingauthor{Labanya K. Guha}
\email{labanya.guha@iiap.res.in}

\begin{abstract}
We present the largest catalogue to date of Galaxies On Top Of Quasars (GOTOQs) -- systems where the sightline to a background quasar passes directly through or very close to a foreground galaxy. Using $\sim$1.1 million quasar spectra from the SDSS-DR16 and DESI Early Data Release, we identify 1,345 unique GOTOQs over $0.0045 \leq z \leq 1.09$, more than quadrupling the number previously known. The catalogue combines both absorption-agnostic and absorption-selected searches, enabling a robust characterisation of gas in galaxies at projected separations $D \lesssim 20$~kpc.  The GOTOQ emission-line ratios indicate that their host galaxies are predominantly normal, star-forming disks with typical dust extinctions of $A_V \simeq 0.5$~mag. We measure the \MgII\ covering fraction at $D \lesssim 20$ kpc and find it to be remarkably high, $f_c = 98.2^{+1.3}_{-4.5}$ per cent, indicating that these systems trace gas-rich, metal-enriched regions at the disk–halo interface. The median colour excess towards the quasar line-of-sight of $E(B-V)=0.087$ is significantly higher than that of typical Mg\,{\sc ii} absorbers, underscoring the dust-rich nature of GOTOQ sightlines.  From a high signal-to-noise composite spectrum, we report the first statistical detection of the diffuse interstellar band at $\lambda4428$ ($W_r = 0.055\pm 0.011$~\AA) and at $\lambda5780$ ($W_r = 0.051\pm 0.012$~\AA), revealing the presence of complex organic molecules at the disk-halo interface. This catalogue provides a powerful reference sample for future multi-wavelength studies of gas flows, cold gas, and dust evolution in the inner circumgalactic medium.
\end{abstract}



\section{Introduction}

Galaxy evolution is governed by a complex interplay between gas accretion, star formation, black hole formation and growth, and a range of feedback processes. A key component in this evolutionary cycle is the circumgalactic medium (CGM), a diffuse halo of gas that surrounds galaxies and typically extends to hundreds of kiloparsecs \citep[e.g.,][]{Tumlinson2017, Peroux2020b}, far beyond the stellar body. Acting as both a reservoir and a conduit for baryons, the CGM regulates the galaxy growth through the exchange of materials between galaxies and their larger-scale environments.

Owing to its low density, the CGM is most effectively studied through quasar absorption-line spectroscopy, which provides sensitive, pencil-beam probes of gas over a wide range of physical conditions and redshifts. Although recent advances in integral field unit (IFU) spectroscopy using instruments such as the Multi Unit Spectroscopic Explorer (MUSE) and the Keck Cosmic Web Imager (KCWI) have enabled direct imaging of the CGM in emission \citep[see, for example,][]{Banerjee2025, Chisholm2020}, quasar absorption lines remain the most powerful and statistically efficient tool for characterizing the CGM, particularly at high redshifts.

Among the various absorption tracers, \MgII\ has been widely used to study the CGM of low-redshift galaxies ($z<1$) from ground-based spectrographs. Over the past two decades, systematic surveys combined with IFU follow-up studies \citep[e.g.,][]{bouche2012, Nielsen_2013, Schroetter2019, Dutta2020, Huang2021, Lundgren2012, Lundgren2021, Guha2022a, Guha2024} have established robust connections between \MgII\ absorbers and their host galaxies or galaxy groups across a broad range of impact parameters ($D$). However, the majority of these investigations are biased toward large separations, typically $D \gtrsim 30$ kpc, leaving a significant gap in our understanding of the inner CGM at $D \lesssim 20$ kpc \citep[see, however,][]{Kacprzak_2013, Guha2022b, Guha2024b}.

Establishing a direct connection between star formation in galaxies and the circumgalactic gas, especially at large impact parameters, remains challenging. On the other hand, examining star-forming galaxies through background quasar sightlines at very low impact parameters -- within regions affected by galactic winds over characteristic star formation timescales -- can offer significant insights into the role of large-scale gas flows in shaping CGM conditions. For example, an outflow moving at 200 $\rm kms^{-1}$ can influence a region up to approximately 20 kpc around galaxies within about 100 Myr, a timescale comparable to that of a typical starburst episode \citep{McQuinn2009}.

A quasar line-of-sight passing near the center of an intervening galaxy (i.e., at ${D \rm \lesssim 20~ kpc}$) is likely to probe the gaseous disk or disk-halo interface of a galaxy, providing the optimal means to link large-scale gas flows with galaxy evolution. Investigating gas flows at these scales is essential for understanding the mechanisms that sustain or quench star formation in galaxies. The most effective strategy for identifying such quasar-galaxy pairs is to search for nebular emission lines from the intervening galaxy in the spectrum of the background quasar. This method has proven highly efficient for identifying host galaxies of \MgII\ absorption systems situated at very small angular separations -- typically at sub-arcsec separations \citep{Noterdaeme_2010, york2012, Straka_2015, Joshi2017, Joshi2018, Kulkarni2022, Rubin2022}. Such systems are referred to as Galaxies On Top of Quasars \citep[GOTOQ;][]{york2012}.

Past studies utilize one of the two primary methods to search for \gotoqs: absorption-agnostic and absorption-selected. The absorption-agnostic method involves detecting multiple strong nebular emission lines from an intervening galaxy in the spectrum of a background quasar without any prior knowledge on the presence of absorption lines towards the quasar line-of-sight \citep[see, for example,][]{Noterdaeme2010, straka2013}. This method is particularly effective at lower redshifts ($z \lesssim 0.4$) because all the strong nebular emission lines in the rest-frame optical wavelengths (e.g., \OII\ $\lambda\lambda$ 3727, 3729,  \OIII\ $\lambda$ 5008, and H$\alpha\, \lambda$ 6564) can be observed using ground-based spectrographs. However, at higher redshifts ($z \gtrsim 0.4$), the H$\alpha$ line gets redshifted beyond the observable range of these spectrographs, reducing the efficiency of the technique. Using this approach, \citet{Straka_2015} compiled a total of 103 such systems from the Sloan Digital Sky Survey (SDSS) DR7, covering the redshift range, $0 < z < 0.84$. 

In the absorption-selected method, one starts from systems identified based on the presence of intervening metal absorption (e.g., \MgII\ $\lambda\lambda$ 2796, 2803) in the quasar spectrum and searches for the presence of associated nebular emissions at the absorption redshift. This method is particularly efficient for identifying \gotoqs\ at higher redshifts, $z \gtrsim 0.4$. Starting from the SDSS DR12 \FeII / \MgII\ metal absorber catalog \citep{Zhu2013}, \citet{Joshi2017} have identified a total of 198 \gotoqs\ over the redshift range, $0.3 < z < 1.1$. As a result, approximately 300 unique \gotoqs\ are currently known in the literature, based on earlier SDSS data releases.

Although detecting these galaxies in \HI\ 21-cm emission at $z\gtrsim 0.2$ is challenging, owing to their distinctive geometrical configuration where the line-of-sight to a background quasar passes either through or very close to the star-forming disk of a foreground galaxy, \gotoqs\ provide an exceptional opportunity to study the cold gas in the disk and their kinematics through \HI\ 21-cm absorption if the background quasar happens to be radio loud.
A recent survey on \gotoqs\ by \citet{Guha2025} reported a staggering $\approx90\%$ detection rate of \HI\ 21-cm absorption, which is significantly higher than that obtained from searches based on metal lines or damped \lya\ absorbers, thereby highlighting the efficiency of \gotoqs\ in tracing cold gas reservoirs. As seen in the Galactic interstellar medium (ISM), GOTOQs also show a correlation between the line-of-sight reddening and the \HI\ 21-cm absorption line optical depth.

Moreover, \gotoqs\ serve as valuable laboratories for investigating large-scale gas flows and the cycling of material through galactic fountains \citep[see, e.g.,][]{Rubin2022} by combining quasar absorption lines with down-the-barrel absorption from galaxies. Observational studies have revealed a possible bimodal azimuthal dependence of metal-line absorption around galaxies, with gas accretion and outflows occurring preferentially along the major and minor axes, respectively \citep[e.g.,][]{kacprzak2012}. However, other studies \citep[e.g.,][]{Pointon2019} have found no such orientation dependence in the CGM metallicity, possibly due to gas mixing that ``wash-out'' the bimodality. A well-curated and statistically significant sample of \gotoqs\ can settle this debate, as the effects of gas mixing are expected to be minimal at the small impact parameters probed by these systems.

In this study, we revisit the search for \gotoqs\ using the latest data from SDSS DR16 \citep{sdss_dr16} and the Early Data Release of the Dark Energy Spectroscopic Instrument \citep[DESI-EDR;][]{desi_edr}. This effort has resulted in the compilation of the largest sample of \gotoqs\ to date, increasing the number known in the literature by more than a factor of four. { Building a large sample of \gotoqs\ is particularly important for probing the CGM at very low impact parameters, a regime that is rarely sampled in conventional CGM surveys. However, the detection of GOTOQs depends on several factors---(i) fiber diameter, (ii) the signal-to-noise ratio and free spectral range of the observed spectrum, (iii) the criteria adopted for identifying detections, and (iv) the relative brightness of the galaxies and quasars---making it difficult to uniquely quantify the selection function \citep[see, e.g.,][]{Noterdaeme_2010}. Consequently, while the catalog is extremely valuable for follow-up studies, it must be used with caution when drawing statistical or physical inferences.}
This paper is organized as follows. In Section \ref{sec:sample}, we describe the methods used to identify \gotoqs\ and provide details of the sample. In Section \ref{sec:analysis_result}, we present our analysis of the \gotoqs\ along with the key results. Section \ref{sec:discussion} summarizes the main findings and discusses their implications. Throughout this work, we adopt a flat $\Lambda$CDM cosmology with $H_0 = 70\,\mathrm{km\,s^{-1}\,Mpc^{-1}}$ and $\Omega_m = 0.3$.

\section{The \gotoqs\ sample}
\label{sec:sample}
Our search for \gotoqs\ involves both the methods invoked in the literature -- absorption-agnostic as well as absorption-selected searches. In the following, we briefly describe the methods employed to search for \gotoqs. 

\subsection{Absorption-agnostic searches of \gotoqs}
\label{sec:agnostic_searches}

\begin{figure*}
    \centering
    \includegraphics[width=0.99\textwidth]{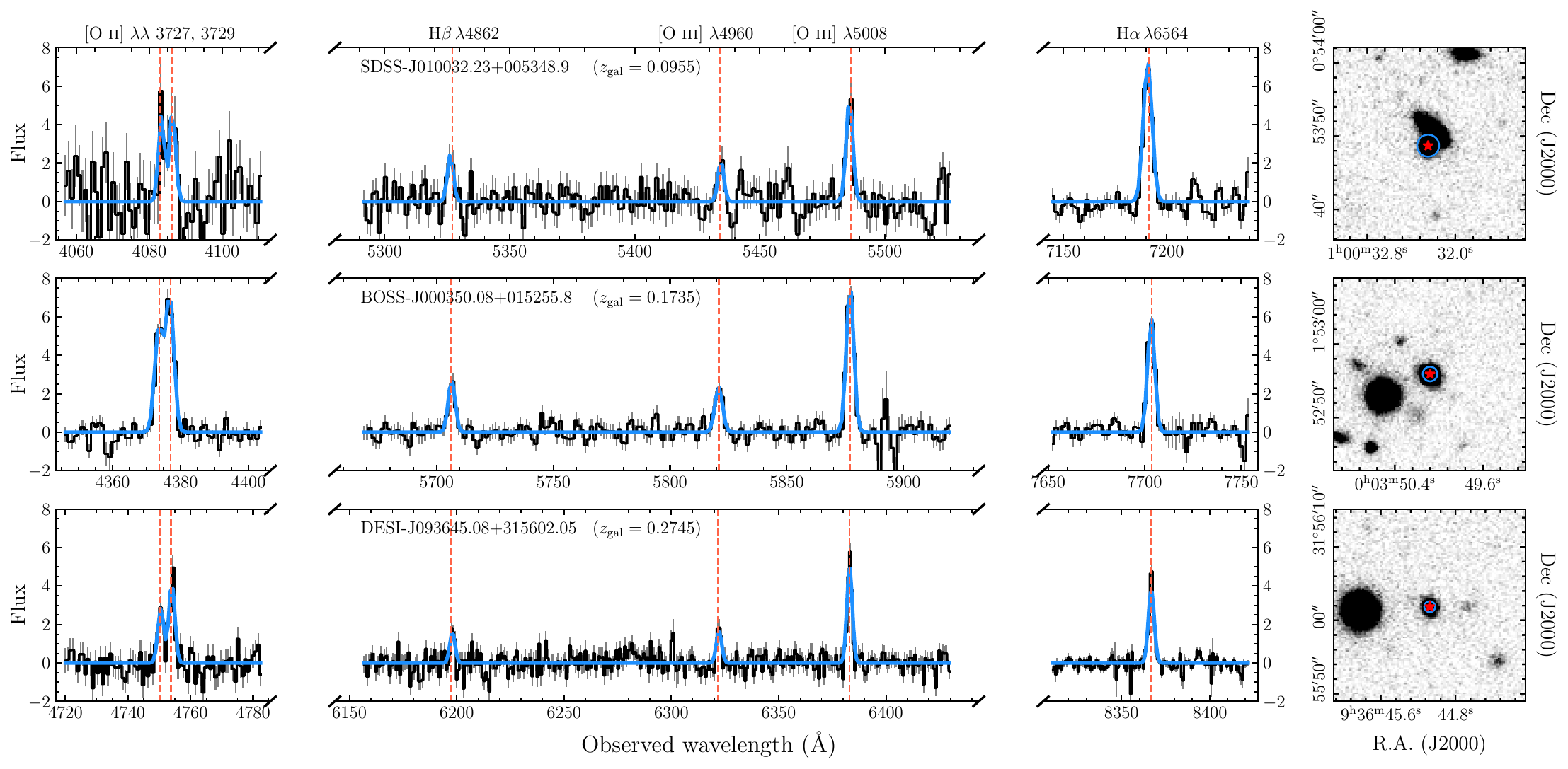}
    \caption{Examples of absorption-agnostic \gotoqs. Each row represents a different quasar observed with SDSS, BOSS, and DESI spectrographs, respectively, from top to bottom. In each horizontal panel: the left plot displays the nebular emission lines detected in the continuum-subtracted quasar spectra with the solid blue line representing the Gaussian fits to the nebular emission lines and the right plot shows a composite $griz$ band DESI-LIS image of the quasar field, with the quasar marked by a red star and the observing fiber outlined by a blue circle. Flux scales in the left panels are in the units of $\rm{ 10^{-17}\, ergs\, s^{-1}\, cm^{-2}\, \text{\AA}^{-1}}$.}
    \label{fig:absorption-agnostic-gotoqs}
\end{figure*}

To search for \gotoqs, without any prior knowledge of the presence of any absorption towards the quasar line-of-sight, we start with a quasar spectrum. As the principle component analysis (PCA) based continuum of the quasar spectrum provided in SDSS is not always adequate for the analysis presented here, we first perform an automated fitting of the quasar continuum spectra, using \textsc{QSmooth} \citep{Durovcikova2020} to obtain a continuum-subtracted quasar spectrum. In brief, \textsc{QSmooth} processes the spectrum by initially computing a running median to capture the continuum and broad emission lines. Subsequently, a peak-finding algorithm identifies peaks above this median. These peaks are then interpolated to create an upper envelope of the spectrum, which is then subtracted. The method then applies RANSAC regression \citep{fischler1981random} to the residuals to eliminate absorption features. The inliers are interpolated and smoothed using a running median, ultimately producing an adequate quasar continuum. We then subtract this continuum to obtain the continuum-subtracted quasar spectrum.

To search for nebular emission lines from a foreground galaxy superimposed on the continuum-subtracted quasar spectrum, we first mask $\rm \pm 3000\, km\, s^{-1}$ around the prominent quasar emission lines to ensure that the identified emission features are not due to the quasar itself. Subsequently, we smooth the continuum-subtracted quasar spectrum by three pixels to enhance the prominence of any potential emission lines from the foreground galaxy. A peak-finding algorithm is then applied on the smoothed spectrum to identify emission features at $2\sigma$ significance and calculate the wavelength ratios for all possible pairs of these features. We then assess how many of these ratios are consistent within a rest-frame velocity offset of $100$ \kms\ with the wavelength ratios of known strong nebular emission lines (\OII\ $\lambda$ 3728, $\rm H\beta\, \lambda$ 4862, \OIII\ $\lambda$ 4960, \OIII\ $\lambda$ 5008, and $\rm H\alpha\, \lambda$ 6564). If there are at least three emission line pairs that match these wavelength ratios, we classify the quasar spectrum as a potential GOTOQ candidate. Finally, each GOTOQ candidate is visually inspected and confirmed to contain at least three nebular emission lines, each detected with a significance greater than or equal to $\rm 3\sigma$, to verify the presence of a foreground galaxy on top of the quasar. 
{ Although, in principle, our search method enables the detection of \gotoqs\ over the full redshift range $z < z_{qso}$, we restrict our analysis to $z < 1$ to ensure that the \OIII\ $\lambda$5008 line remains within the observable spectral window, thereby guaranteeing coverage of at least three prominent emission lines.}

To search for \gotoqs, we apply our search pipeline to the quasar spectra available in the SDSS-DR16 and DESI-EDR surveys. The SDSS-DR16 quasar catalog \citep{Lyke2020} includes approximately 0.75 million unique quasars, encompassing nearly 1 million observed spectra across various epochs. The DESI-EDR survey, on the other hand, covering only 2\% of the DESI sky, contains about 95000 quasar spectra. Therefore, we search for \gotoqs\ in almost 1.1 million quasar spectra towards more than 0.8 million quasars. Our search identified 681 \gotoqs\ from SDSS-DR16 and 56 \gotoqs\ from the DESI-EDR survey. With 6 overlaps, the combined sample comprises 731 unique \gotoqs\ spanning the redshift range, $0.0045 \leqslant z \leqslant 0.9782$ with a median value of 0.33. In Figure \ref{fig:absorption-agnostic-gotoqs}, we show three examples of absorption-agnostic \gotoqs. Each row in this figure represents a different quasar observed with SDSS (fiber diameter of $3^{\prime\prime}$), BOSS (fiber diameter of $2^{\prime\prime}$), and DESI (fiber diameter of $1.5^{\prime\prime}$) spectrographs, respectively, from top to bottom. In each horizontal panel, the left plot displays the nebular emission lines detected in the continuum-subtracted quasar spectra and the right plot shows a composite $griz$ band DESI-LIS image of the quasar field, with the quasar marked by a red star and the observing fiber outlined by a blue circle.


\subsection{Absorption-selected searches of \gotoqs}
\label{sec:absorption_selected_searches}

\begin{figure*}
    \centering
    \includegraphics[width=0.99\textwidth]{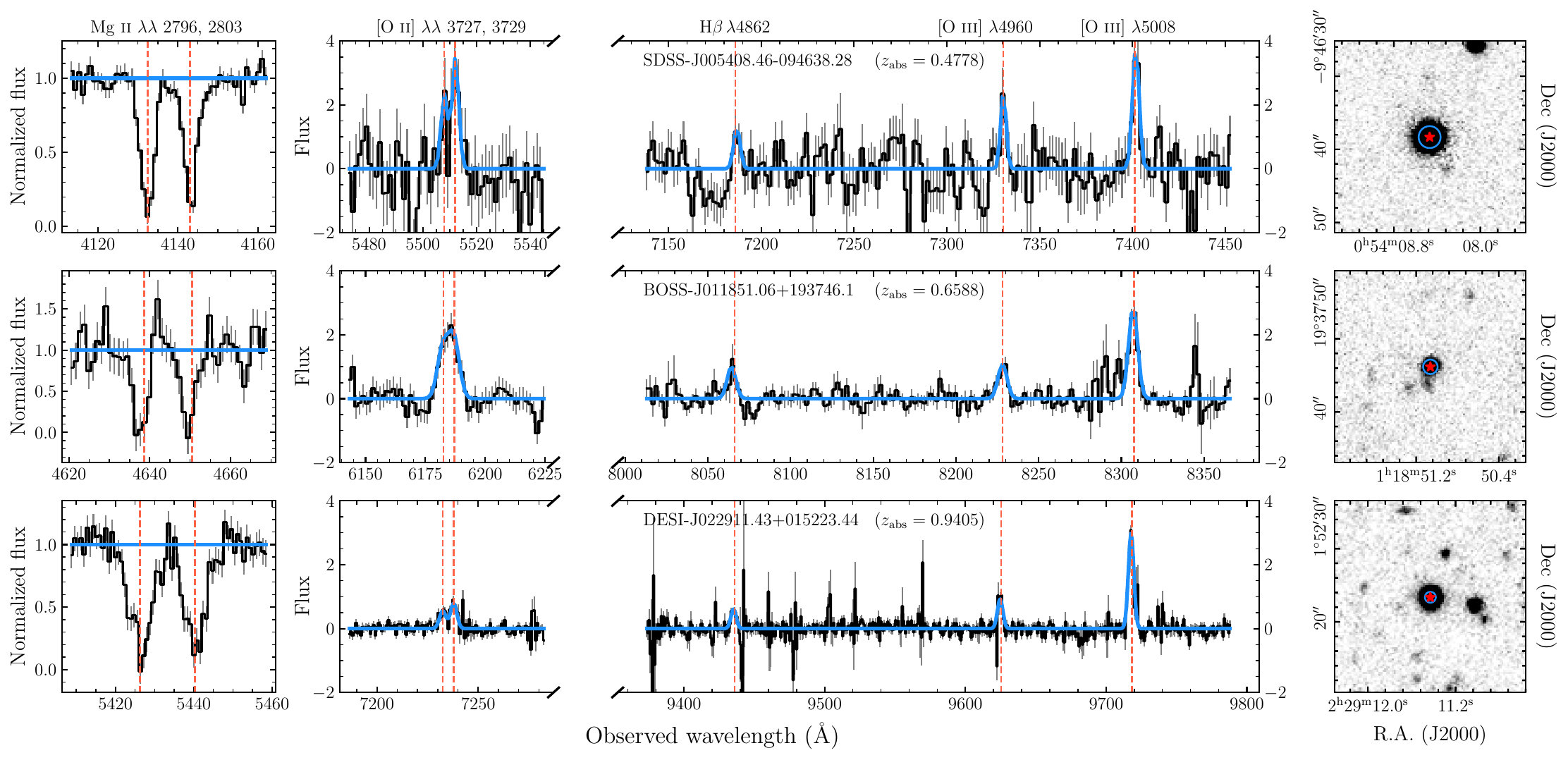}
    \caption{Examples of absorption-selected \gotoqs. Each row represents a different quasar observed with SDSS, BOSS, and DESI spectrographs, respectively, from top to bottom. In each horizontal panel: the left plot displays the intervening \MgII\ absorption in a continuum-normalized spectrum; the middle plot highlights the nebular emission lines detected in the continuum-subtracted quasar spectra, corresponding to the \MgII\ absorption with the solid blue line representing the Gaussian fits to the nebular emission lines; the right plot shows a composite $griz$ band DESI-LIS image of the quasar field, with the quasar marked by a red star and the observing fiber outlined by a blue circle. Flux scales in the middle panels are in the units of $\rm{ 10^{-17}\, ergs\, s^{-1}\, cm^{-2}\, \text{\AA}^{-1}}$.}
    \label{fig:absorption-selected-gotoqs}
\end{figure*}

The efficiency of the absorption-agnostic searches reduces beyond $z \gtrsim 0.4$ as $\rm{H\alpha}$, the most prominent nebular emission line in the optical band, gets redshifted beyond the observable window of ground-based spectrographs. {
In addition, a significant fraction of the spectral region containing the H$\beta$ and [O~{\sc iii}] emission lines for systems at $z>0.4$ is affected by atmospheric absorption and strong sky-line residuals.}
Fortunately, at $z>0.4$, \MgII\ $\rm {\lambda\lambda}\, 2796, 2803$ gets redshifted within the observable window of ground-based telescopes.  The \MgII\ $\rm {\lambda\lambda}\, 2796, 2803$ absorption lines found in the spectra of background quasars, originating from the halos of foreground galaxies, are excellent tracers of cold gas ($T\, \sim 10^4$ K) with a near-unity covering fraction within $D \lesssim 10$ kpc of star-forming galaxies \citep{Anand2021}.

\begin{figure}
    \centering
    \includegraphics[width=0.49\textwidth]{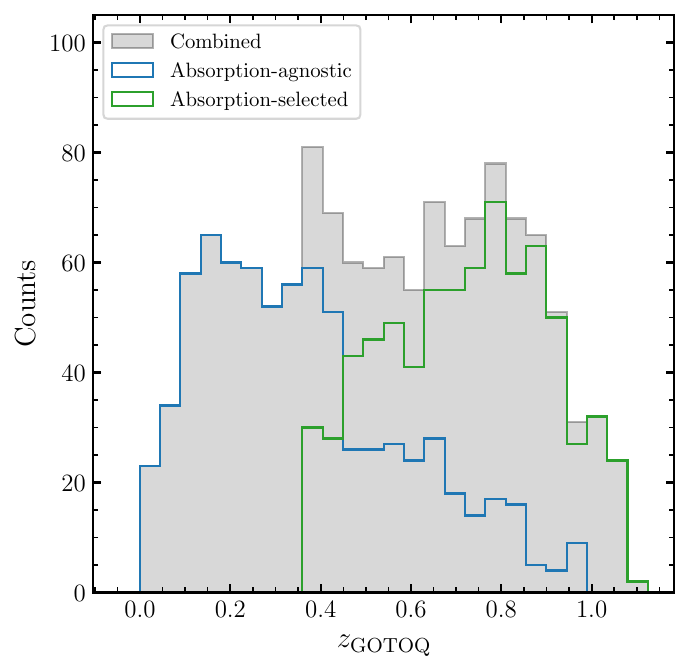}
    \caption{The redshift distribution of the absorption-agnostic (blue histogram) and absorption-selected (green histogram) \gotoqs. The distribution for the combined sample is depicted in gray.}
    \label{fig:zabs_dist}
\end{figure}

To search for \gotoqs\ using this method, we begin with a quasar spectrum that contains an intervening \MgII\ absorption. We then obtain the continuum-subtracted spectrum as described in the first part of Section \ref{sec:agnostic_searches}. In this spectrum, we look for \OII, \OIII, and $\rm{H\beta}$ nebular emission lines associated with \MgII\ absorptions within $\rm \pm500\, km\, s^{-1}$  based on the following criteria: either the \OII\ $\rm \lambda\lambda\, 3726, 3729$ emission is detected with at least $\rm 4\sigma$ significance, or the \OIII\ $\rm \lambda\lambda\, 4960, 5008$ doublets are individually detected with at least $3\sigma$ significance, or both the \OII\ $\rm \lambda\lambda\, 3726, 3729$ and \OIII\ $\lambda\, 5008$ lines are individually detected with at least $3\sigma$ significance. Finally, the spectrum of each selected candidate is visually inspected to confirm it as a GOTOQ system. While our method theoretically allows for the search for \gotoqs\ over the redshift range $0.4 \lesssim z \lesssim 1.7$, ensuring that both \MgII\ absorption and the associated \OII\ emission fall within the observable window of the SDSS and DESI spectrographs, we follow \citet{Joshi2017} and restrict our searches to $0.36 \leqslant z \leqslant 1.1$. This limitation is due to the presence of strong skylines and fringing at higher wavelengths, which are likely to result in a high fraction of false positives.

In the publicly available SDSS \MgII\ absorption catalogs \citep{Anand2021, Zhu2013} up to DR16, we search for \MgII\ absorbers within the redshift range, \zabs\ $\leqslant 1.1$. To minimize false detections, we include only those systems where \MgII\ $\lambda\, 2796$ lines are detected with a significance of at least $\rm 2\sigma$. This criterion results in approximately 88,000 unique \MgII\ absorption systems towards about 75,000 quasars. Similarly, using the same criterion, the DESI-EDR \MgII\ absorption catalog \citep{Napolitano2023} yields approximately 6,200 \MgII\ absorption systems towards about 5,500 quasars. Therefore, we search for nebular emissions associated with about 94,200 \MgII\ absorption systems towards about 80,500 quasars. This resulted in 651 \gotoqs\ in SDSS and 97 \gotoqs\ in DESI-EDR. With 15 overlaps, there are 733 unique \gotoqs\ identified based on the presence of \MgII\ absorption towards the quasar line-of-sight spanning the redshift range, $0.3626 \leqslant z \leqslant 1.0866$ with a median value of 0.73. For about 85\% of the cases, the \gotoqs\ are selected purely based on the \OII\ emission line. In Figure \ref{fig:absorption-selected-gotoqs}, we show three examples of absorption-selected \gotoqs. In this figure, each row represents a different quasar observed with SDSS, BOSS, and DESI spectrographs, respectively, from top to bottom. In each horizontal panel, the left plot displays the intervening \MgII\ absorption in a continuum-normalized spectrum, the middle plot highlights the nebular emission lines detected in the continuum-subtracted quasar spectra, corresponding to the \MgII\ absorption, and the right plot shows a composite $griz$ band DESI-LIS image of the quasar field, with the quasar marked by a red star and the observing fiber outlined by a blue circle.

\subsection{Final Combined Sample}

By merging our absorption-agnostic and absorption-selected searches, we identified 119 overlapping cases. Consequently, our final sample includes 1345 unique \gotoqs\ within the redshift range of $0.0045 \leqslant z \leqslant 1.0866$, with a median redshift of 0.54, sourced from the SDSS-DR16 and DESI-EDR surveys, with typical redshift uncertainty of $\rm 70\, km\, s^{-1}$. This constitutes by far the largest sample of \gotoqs\ ever compiled, more than quadrupling the number of previously known \gotoqs. In Figure \ref{fig:zabs_dist}, we show the redshift distribution of the absorption-agnostic (blue histogram) and absorption-selected (green histogram) \gotoqs. The distribution for the combined sample is depicted in gray. It is evident from the figure that absorption-agnostic searches are more effective at identifying low-redshift \gotoqs\ ($z \lesssim 0.4$), whereas absorption-selected searches dominate the discovery of higher-redshift systems ($0.4 \lesssim z \lesssim 1$). { As discussed earlier, this is partly due to the stringent criteria adopted in our absorption-agnostic search, as well as contamination from atmospheric absorption and sky-line residuals.}

{Table~\ref{tab:sample_summary} summarizes the number of \gotoqs\ identified from each spectrograph and their corresponding sample properties. We identify 140 \gotoqs\ from DESI-EDR, while the BOSS and SDSS spectrographs contribute 1016 and 289 \gotoqs, respectively. Since 82 systems are detected in both the BOSS and SDSS spectroscopic datasets, the combined SDSS-DR16 sample comprises 1223 unique \gotoqs. Furthermore, 18 of these systems are also present in DESI-EDR. After accounting for these duplicates, our final catalog contains a total of 1345 unique \gotoqs.}

\begin{table*}[]
    \centering
    \begin{tabular}{lccc}
    \hline
        Survey  & $N_{\rm GOTOQs}$  & $z$ range  & $L_{\rm [O\, II]}$ range  \\
                &                   &            &          ($\rm erg\, s^{-1}$) \\
         (1)     &    (2)            &   (3)      &     (4)       \\
       \hline
        DESI-EDR  & 140  & $0.0595-1.0849$  &   $7.90\times10^{38}-4.74\times10^{41}$\\
        BOSS      & 1016 & $0.0045-1.0866$  &   $7.11\times10^{36}-1.89\times10^{42}$\\
        SDSS      & 289  & $0.0206-1.0628$  &   $2.48\times10^{38}-2.06\times10^{42}$\\
        \hline
    \end{tabular}
    \caption{Summary of the final combined \gotoqs\ sample. Column (1) lists the survey or instrument from which the \gotoqs\ were identified, Column (2) gives the number of \gotoqs\ contributed by each survey, Column (3) provides the corresponding redshift range of the \gotoqs, and Column (4) lists the range of fiber-measured \OII\ emission-line luminosities for the \gotoqs.}
    \label{tab:sample_summary}
\end{table*}

\section{Analysis \& Results}
\label{sec:analysis_result}

For each identified GOTOQ, we estimate the emission line parameters by fitting the detected nebular lines using Gaussian profiles. Since the \OII\ doublet is often unresolved in low-resolution spectrographs like SDSS, we model the \OII\ emission with a double Gaussian, tying the redshifts and linewidths while allowing the line ratio to vary within the range of $0.34-1.5$ \citep{OsterbrockFerland2006}. The $\rm H\beta$ $\lambda$ 4862, \OIII\ $\lambda\lambda$ 4960, 5008, and $\rm H\alpha$ $\lambda$ 6564 lines are each modeled with a single Gaussian. In Figures \ref{fig:absorption-agnostic-gotoqs} and \ref{fig:absorption-selected-gotoqs}, we also present the best-fitting Gaussian profiles (shown in blue) for the nebular emission lines from foreground galaxies present on top of the background quasar spectra. The integrated fluxes for the nebular lines are then measured based on the fitted Gaussian parameters.
In the following, we examine the emission line properties of GOTOQs and compare them with those derived from absorption lines, as well as with the properties of field galaxies.

\subsection{The \OII\  Line Luminosity: Fiber Effects}
\label{subsec:oii_luminosity}
\begin{figure}
    \centering
    \includegraphics[width=0.5\textwidth]{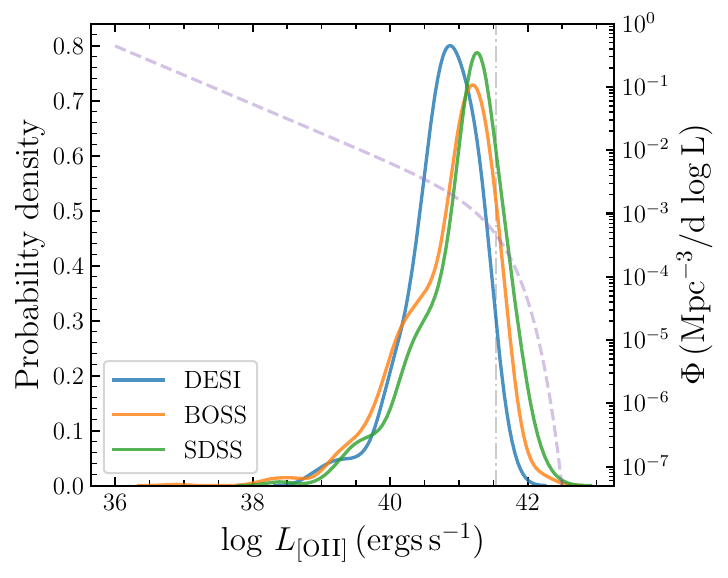}
    \caption{The kernel density estimations of \OII\ line luminosities (considering only detections and excluding upper limits) for the \gotoqs\ in our sample is shown from DESI (blue), BOSS (orange), and SDSS (green). For comparison, the \OII\ luminosity function of field galaxies at the median redshift of our GOTOQ sample, $z = 0.54$, is shown as a dashed purple curve. The brown vertical dot-dashed line marks the characteristic \OII\ luminosity ($L^\star_{\rm [O\, II]}$) at $z = 0.54$ \citep{Comparat2016}.}
    \label{fig:o2lum_hist}
\end{figure}

In Figure \ref{fig:o2lum_hist}, we show the distribution of the measured \OII\ line luminosities captured within the fiber (considering only detections and excluding upper limits) for the \gotoqs\ in our combined sample from DESI (blue), BOSS (orange), and SDSS (green). For comparison, the \OII\ luminosity function of field galaxies at the median redshift of our GOTOQ sample, $z = 0.54$, is shown as a dashed purple curve. The brown vertical dot-dashed line marks the characteristic \OII\ luminosity at $z = 0.54$ \citep{Comparat2016}. The \OII\ line luminosity for the clear detections in the final combined sample of \gotoqs\ varies from $7.11\times10^{36}\, \rm erg\, s^{-1}$ to $2.06\times10^{42}\, \rm erg\, s^{-1}$ with a median value of $1.04\times10^{41}\, \rm erg\, s^{-1}$. In cases where the same GOTOQ was observed with multiple spectrographs, we include the measured line luminosities from all instruments. If a GOTOQ was observed multiple times with the same spectrograph, we consider the \OII\ line luminosity from the co-added spectra. However, line luminosities of GOTOQs in general are severely affected by the fiber effects and dust attenuation, and thus what we measure should be considered as the lower limits.

\begin{figure}
    \centering
    \includegraphics[width=0.5\textwidth]{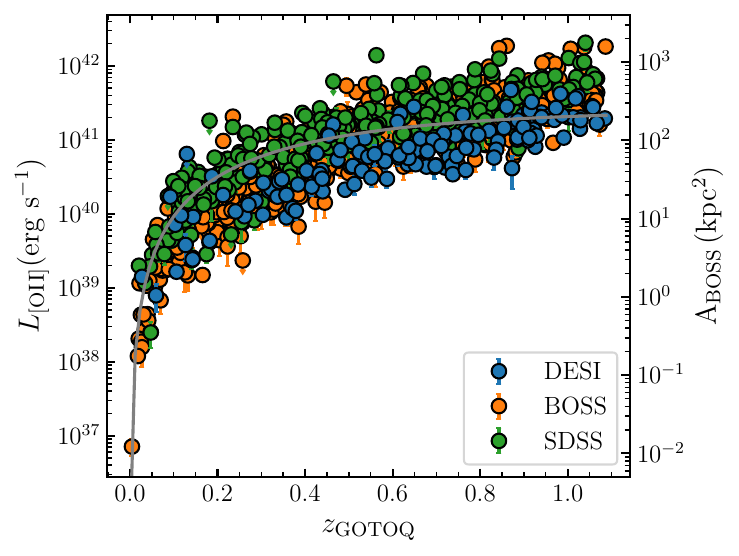}
    \caption{The \OII\ line luminosities of the GOTOQs captured within the fibers of DESI (blue) SDSS (green) and BOSS (orange) spectrographs are plotted against their redshift. The solid gray line corresponds to the physical area a BOSS fiber covers as a function of redshift.}
    \label{fig:z_vs_o2lum}
\end{figure}

To illustrate the effects of fiber size, Figure \ref{fig:z_vs_o2lum} shows the \OII\ line luminosities of the \gotoqs\ as a function of redshift. Regardless of fiber size, a strong correlation is observed between the measured \OII\ luminosities and the redshifts of the \gotoqs. A Kendall-$\tau$ rank correlation test, including upper limits \citep{Isobe1986}, yields a correlation coefficient of $\tau_r = 0.677$ with a $p$-value essentially zero, indicating a statistically strong and significant trend. However, this apparent evolution in the \OII\ line luminosity is not intrinsic, but rather a consequence of the increasing projected area of the fiber with redshift (in addition to typical trend expected for a flux limited sample). The solid gray line in Figure \ref{fig:z_vs_o2lum} illustrates this evolution, showing that the increase in \OII\ luminosity basically tracks the growth of the projected area within the fiber.

\subsection{Nebular Line Ratios}
\label{subsec:nebular_line_ratios}

\begin{figure}
    \centering
    \includegraphics[width=0.5\textwidth]{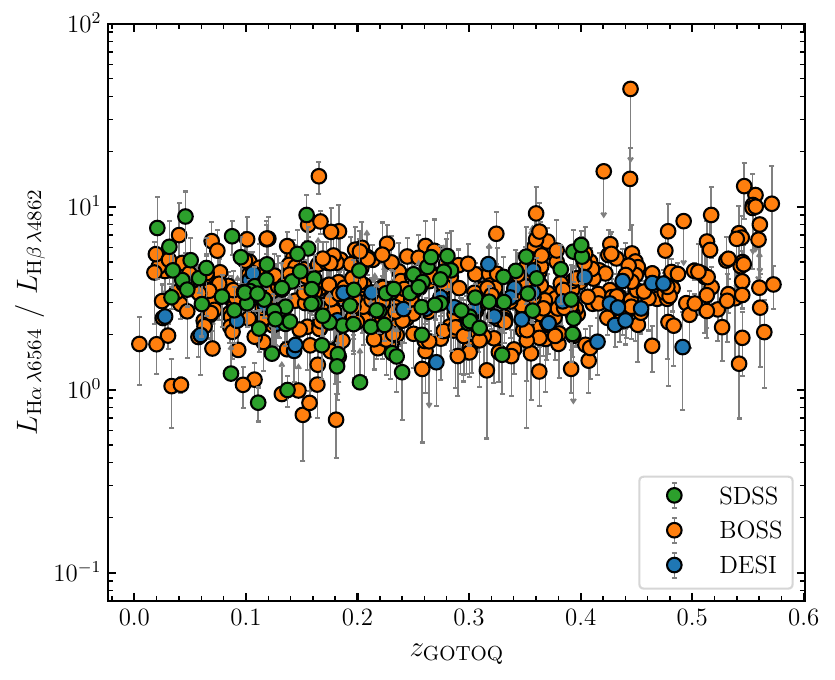}
    \caption{The redshift evolution of the $\rm H\alpha / H\beta$ line ratios for the \gotoqs\ in our sample, where $\rm H\alpha$ line is covered.}
    \label{fig:ha_hb_lineratios}
\end{figure}
While the measured line luminosities of the \gotoqs\ are severely affected by the fiber effects, nebular line ratios, in principle, characterizes the \gotoqs\ better as the line ratios are expected to be least affected by the fiber effects.

\subsubsection{$\rm H\alpha / H\beta$ Line Ratio}
\label{sec:ha_hb_ratio}
The $\rm H\alpha / H\beta$ line ratio is sensitive to dust and serves as a diagnostic of dust attenuation in the interstellar medium (particularly in newly star-forming regions) of galaxies. In the absence of dust, this ratio is expected to be approximately 2.86. However, in the presence of dust, a lower $\rm H\beta / H\alpha$ is observed as the dust attenuates the $\rm H\beta$ line more than the $\rm H\alpha$ line -- a phenomenon known as the Balmer decrement \citep{OsterbrockFerland2006}. In Figure \ref{fig:ha_hb_lineratios}, we present the redshift evolution of the $\rm H\alpha / H\beta$ line ratios for the \gotoqs\ in our sample where the $\rm H\alpha$ line is covered, i.e, $z \lesssim 0.5$. A Kendall $\tau$ rank correlation test between $z_{\rm GOTOQ}$ and the $\rm H\alpha / H\beta$ ratio, accounting for both upper and lower limits, yields a correlation coefficient of $\tau_r = 0.02$ with a $p$-value of 0.58, indicating no redshift evolution of the ratio over the redshift range, $0 < z \lesssim 0.5$. Excluding the lower and the upper limits, the median value of $\rm H\alpha/H\beta$ line ratio for the \gotoqs\ is $3.33^{+1.57}_{-1.09}$, with the errors corresponding to the 34 percentiles from the median. Assuming the \citet{Calzetti2000} extinction curve, this corresponds to the median dust extinction coefficient of $A_{\rm V} \approx 0.53^{+1.33}_{-1.38}$ mag. 

\subsubsection{\OIII\ / \OII\ Line Ratio}

\begin{figure}
    \centering
    \includegraphics[width=0.5\textwidth]{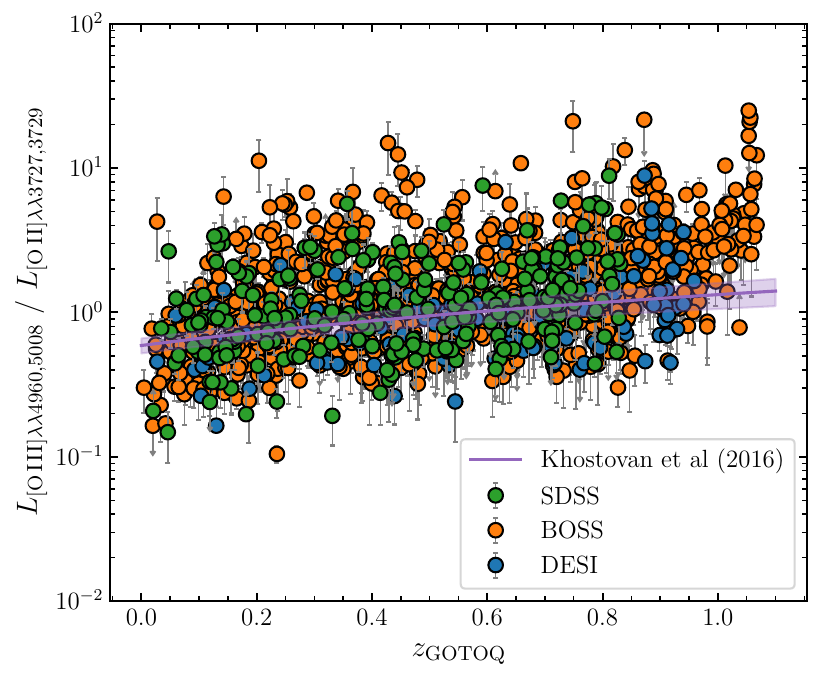}
    \caption{The \OIII\ / \OII\ line ratios for the \gotoqs\ in our sample. The green, orange, and blue points correspond to the GOTOQs detected with SDSS, BOSS, and DESI spectrographs respectively. The redshift evolution of the \OIII\ / \OII\ ratio for the field galaxies, as modeled by a power-law fit from \citet{Khostovan2016}, is represented by the solid magenta line, with the shaded region indicating the 1$\sigma$ uncertainty.}
    \label{fig:o3_o2_lineratios}
\end{figure}

The \OIII\ / \OII\ line ratio is sensitive to the hardness of the ionizing radiation field. Observational studies have consistently reported an increase in this ratio for star-forming galaxies over the redshift range $z = 0-5$. This trend suggests that high-redshift galaxies typically exhibit higher ionization parameters, lower metallicities, harder stellar ionizing spectra, and elevated electron densities compared to their local counterparts \citep{Kewley2002}.

In Figure \ref{fig:o3_o2_lineratios}, we show the \OIII\ $\lambda\lambda$4960, 5008 / \OII\ $\lambda\lambda$3727, 3729 line ratios of the GOTOQs in our sample as a function of redshift. Since \OIII\ $\lambda$4960 emission is not detected most of the times, we scale the \OIII\ $\lambda$5008 luminosity by a factor of 4/3, since the $\lambda$5008 line is expected to be three times stronger than the $\lambda$4960 line \citep{Galavis1997}. The green, orange, and blue points indicate GOTOQs detected with SDSS, BOSS, and DESI spectra, respectively. For comparison, the redshift evolution of the \OIII/\OII\ ratio for field galaxies, modeled with a power-law fit by \citet{Khostovan2016}, is shown as the solid magenta line, with the shaded region marking the 1$\sigma$ uncertainty. The similarity between the GOTOQ measurements and the field-galaxy trend demonstrates that GOTOQs predominantly select normal star-forming field galaxies, rather than a distinct or unusual population. 
It is worth noting that, unlike in Figure \ref{fig:z_vs_o2lum}, where the measured line luminosities at a given redshift are systematically higher for larger fibers and there is minimal overlap between the SDSS and DESI measurements, Figures \ref{fig:ha_hb_lineratios} and \ref{fig:o3_o2_lineratios} show that the line ratios span overlapping regions across different spectrographs. This indicates that, in contrast to absolute line luminosities, line ratios are largely insensitive to fiber size effects.

\subsection{Absorption-selected \gotoqs : The $\rm W_{2796}$ Distribution}
The \MgII\ $\lambda$2796 rest-frame equivalent widths ($W_{2796}$) of the \gotoqs\ in our absorption-selected sample range from 0.31\,\AA\ to 5.90\,\AA, with a median value of 2.16\,\AA. In comparison, the \MgII\ absorption systems compiled from SDSS-DR16 and DESI-EDR, over the parent sample from which the absorption-selected searches for \gotoqs\ were performed, has a median $W_{2796}$ of 1.16\,\AA. Figure \ref{fig:width_dist} presents the normalized distributions of $W_{2796}$, showing the SDSS and DESI-EDR absorbers in blue and the \gotoqs\ in orange. It is evident from the figure, that the $W_{2796}$ tends to be systematically higher when nebular emission lines are detected. A Kolmogorov-Smirnov (KS) test between the $W_{2796}$ distributions of the absorption-selected \gotoqs\ and the parent \MgII\ absorber population confirms that the two are statistically distinct (with a $p$-value almost zero). { We confirm this result even when we consider SDSS or DESI-EDR systems separately.}
This is in line with the earlier finding of \citet{Joshi2017}, who showed the distribution of $W_{2796}$ in GOTOQs is consistent with what has been seen for Mikly Way sightlines.

\subsubsection{GOTOQ detection rate as a function of $W_{2796}$}

\begin{figure}
    \centering
    \includegraphics[width=\linewidth]{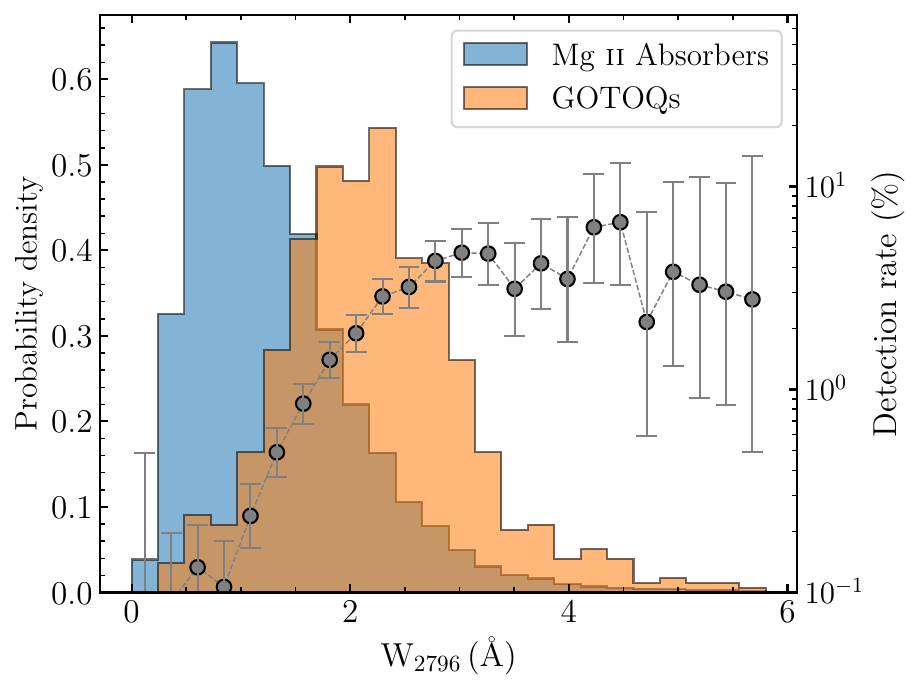}
    \caption{Normalized rest-frame equivalent width distributions of \MgII\ absorption systems from the SDSS and DESI-EDR catalogs (blue), compared with those for the \gotoqs\ (orange). The absorption-selected \gotoqs\ were specifically searched for corresponding systems in these catalogs and the gray circles correspond to the fraction of \gotoqs\ in each rest equivalent width bin. The error bars on the detection rates corresponds to the 95\% Wilson confidence intervals.}
    \label{fig:width_dist}
\end{figure}

Figure~\ref{fig:width_dist} also shows the detection rate of nebular emission as a function of $W_{2796}$ (gray circles). The \gotoqs\ detection fraction increases steadily up to $W_{2796} \approx 3$~\AA, where it reaches a peak of roughly 5 per cent, and then remains nearly flat at larger equivalent widths. If one were to rely solely on the empirical anti-correlation between $W_{2796}$ and impact parameter \citep{Chen_2010}, values of $W_{2796} > 3$~\AA\ would typically correspond to $D \le 5$~kpc. In such cases, the \OII\ emission from the host galaxy should fall well within the spectroscopic fibre. The persistently low detection rate of \OII\ emission, even among systems with very strong \MgII\ absorption, therefore likely reflects several contributing factors: (i) significant scatter in the $W_{2796}$–$D$ relation, as many high-$W_{2796}$ systems may arise in non-isolated or group environments \citep[see][]{Guha2022a, Guha2024}; (ii) a non-negligible contribution from passive galaxies that exhibit large $W_{2796}$ but weak or no nebular emission; and (iii) observational biases that hinder the detection of faint emission lines superimposed on quasar continua, as discussed by \citet{Noterdaeme_2010}. We return to the $W_{2796}$–$D$ relation for our GOTOQ sample in Section~\ref{sec:WvsD}.

\subsubsection{Correlation between $W_{2796}$ and $L_{[\text{O\, II]}}$}
We now turn to the correlation between the \OII\ line luminosity of the absorption-selected \gotoqs\ and the $W_{2796}$ measured along the quasar sightline. A strong correlation between these two quantities has previously been reported in stacking analyses of \MgII\ absorption systems \citep{Noterdaeme2010, Menard2011}. \citet{Joshi2017} demonstrated that such a correlation could arise from the nebular emission line detection rate being dependent on $W_{2796}$ and not necessarily due to a physical connection between star-formation rate (as probed by \OII\ line) and gas flows (as probed by $W_{2796}$). They found a large scatter in $L_{[\text{O\, II]}}$ for a given  $W_{2796}$ and vice versa and reported a weak $3\sigma$ correlation between the two quantities.

\begin{figure}
    \centering
    \includegraphics[width=\linewidth]{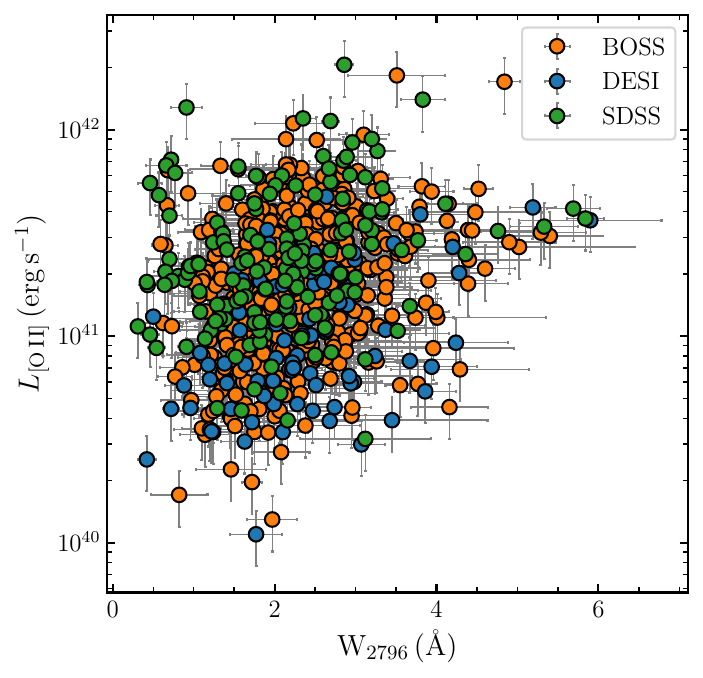}
    \caption{The \OII\ line luminosity of the \gotoqs\ versus the rest frame equivalent width of the \MgII\ $\lambda$ 2796 line for the absorption-selected sample. The blue, green, and orange points correspond to the \gotoqs\ observed with DESI, SDSS, and BOSS spectrographs, respectively.
    }
    \label{fig:width_vs_o2lum}
\end{figure}

In Figure~\ref{fig:width_vs_o2lum}, we plot the \OII\ line luminosities of the \gotoqs\ against the rest-frame equivalent widths of the \MgII\ $\lambda2796$ line, with points color-coded by spectrograph. At fixed $W_{2796}$, the observed $L_{[\text{O II}]}$ spans a wide range, and conversely a given $L_{[\text{O II}]}$ corresponds to a broad distribution of $W_{2796}$. Owing to fiber losses, the \OII\ luminosities measured from SDSS and DESI spectra show little overlap, illustrating how aperture effects can strongly bias the observed luminosity distribution. Despite this, among the direct detections we find a weak but statistically significant positive correlation between $L_{[\text{O II}]}$ and $W_{2796}$, with a Spearman rank coefficient of $r_S = 0.25$ and a near-zero $p$-value. The trend is largely driven by the scarcity of low \OII-luminosity systems at high equivalent widths. {While this could very well be driven by fiber-related effects, if true this correlation can provide important physical insight.}

{For example,} Since $W_{2796}$ primarily traces the number of absorbing components and their velocity dispersion rather than column density, it is natural to speculate that the largest equivalent widths arise in systems hosting star-formation-driven outflows, whose strength correlates with nebular emission-line luminosity. However, recent work by \citet{Fernandez2025}, comparing down-the-barrel absorption with quasar sightlines in strong \MgII\ absorbers, suggests that a significant fraction of the absorbing components may instead be associated with infalling gas. Such infall signatures are rarely seen in field galaxies \citep[e.g.][]{Rubin2014}. A larger, statistically uniform sample of strong \MgII\ absorbers with both emission- and absorption-line information will therefore be required to disentangle the relative roles of inflows and outflows in producing large $W_{2796}$ values.

{An alternative explanation could arise from the strong anti-correlation between $W_{2796}$ and impact parameter ($D$) seen in the general \MgII\ absorber population (coupled with the fiber loss issues discussed above): high-$W_{2796}$ systems would, on average, have smaller $D$ and therefore suffer less fiber loss, leading to a correlation between $W_{2796}$ and $L_{[\text{O\, II]}}$. In the following section, however, we demonstrate that this $W_{2796}$–$D$ anti-correlation flattens out for \gotoqs, implying that the observed $W_{2796}$–$L_{[\text{O\,II]}}$ correlation may not be driven solely by impact-parameter effects.}

\subsection{Absorption-selected \gotoqs : $\rm W_{2796}$ versus Impact Parameters}
\label{sec:WvsD}
Through spectroscopic follow-up of \gotoqs, \citet{Guha2024b} demonstrated that, in many cases, the impact parameters of \gotoqs\ can be reliably determined directly from broadband photometry. They showed that GOTOQs with extended photometric counterparts \citep[see Figure 1 of][]{Guha2022b} whose photometric redshifts agree within $2\sigma$ of the \MgII\ absorption redshifts are indeed the galaxies responsible for the absorption. Using this approach, impact parameters were determined for 74 \gotoqs\ \citep{Guha2022b} out of a parent sample of 198 compiled by \citet{Joshi2017}. In this work, we adopt the same method to identify foreground galaxies and measure their impact parameters.

Based on visual inspection of the absorption-selected \gotoqs\ in our absorption-centric sample using DESI-LIS images, we identified the foreground galaxies and measured their impact parameters from photometric redshifts for 267 systems, corresponding to about 36.5\% of the sample. {This identification rate is broadly consistent across the different spectroscopic datasets and does not vary significantly between the SDSS, BOSS, and DESI subsamples}. The derived impact parameters range from 3.8 to 23.6 kpc, with a median value of 11.6 kpc. A Spearman rank correlation test between $W_{2796}$ and $D$ for the \gotoqs\ yields a rank correlation coefficient of $r_S = -0.05$ with a $p$-value of 0.43, indicating no significant (anti-)correlation. This confirms the flattening of $W_{2796}$ at small $D$, as reported in previous studies \citep{Kacprzak_2013, Guha2022b}.

\begin{figure}
    \centering
    \includegraphics[width=\linewidth]{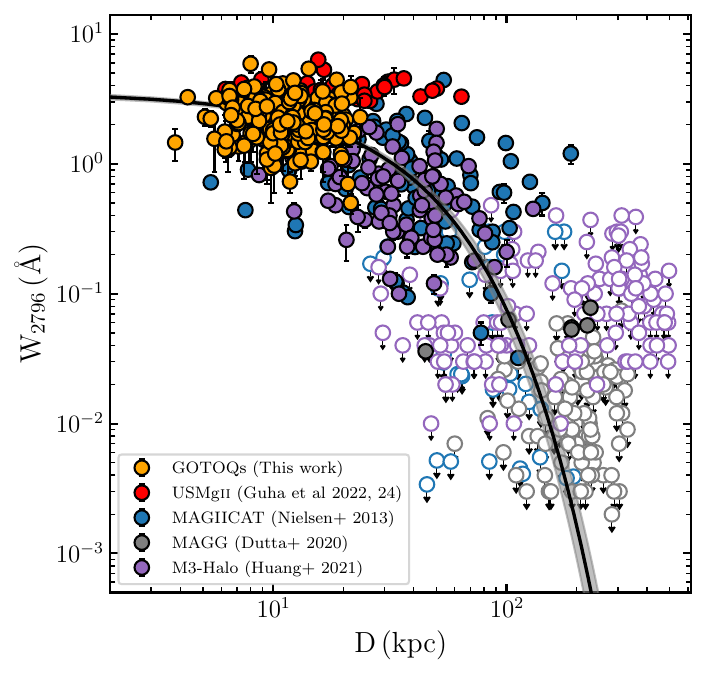}
    \caption{$W_{2796}$–$D$ anti-correlation for isolated galaxies. Orange points show the GOTOQs with photometric extensions from this work, red points represent the \usmg\ absorbers \citep{Guha2022a, Guha2024}, blue points are from the \textsc{MAGiiCAT} survey \citep{Nielsen_2013}, gray points are from the \textsc{MAGG} survey \citep{Dutta2020}, and the purple points are from \citet{Huang2021}. Filled circles denote detections, while open circles indicate upper limits in $W_{2796}$ from non-detections. The best-fit log-linear relation is shown as a solid black line, with the shaded gray band representing the $1\sigma$ uncertainty. The GOTOQ sample densely populates the previously sparsely sampled regime at $D \lesssim 20$ kpc}
    \label{fig:width_vs_rho_all}
\end{figure}

In Figure \ref{fig:width_vs_rho_all}, orange points represent the GOTOQs with photometric extensions from this work, red points correspond to the \usmg\ absorbers \citep{Guha2022a, Guha2024}, blue points are from the \textsc{MAGiiCAT} survey \citep{Nielsen_2013}, gray points are from the \textsc{MAGG} survey \citep{Dutta2020}, and the purple points are from \citet{Huang2021}. Filled circles denote detections, while open circles represent $3\sigma$ upper limits from non-detections.

{The observed $\rm W_{2796}$--$D$ anti-correlation has traditionally been characterized using either a power-law relation \citep[e.g.,][]{Huang2021} or a log-linear fit \citep[e.g.,][]{Nielsen_2013}. However, studies focusing on low impact parameter systems \citep{Kacprzak_2013} and exceptionally strong absorbers such as \usmg\ systems \citep{Guha2022a, Guha2024} have shown that this anti-correlation flattens at small impact parameters ($D \lesssim 20$ kpc). Moreover, \citet{Guha2024} demonstrated through a non-parametric rank-correlation analysis that the anti-correlation effectively disappears for absorbers with $W_{2796} > 1.5$\AA. Given that \gotoqs\ predominantly probe very small impact parameters and are associated with strong \MgII\ absorption, these systems lie precisely in the regime where the conventional $\rm W_{2796}$--$D$ anti-correlation is expected to weaken or vanish. Consistent with this picture, our non-parametric rank-correlation analysis does not reveal any statistically significant anti-correlation. We therefore adopt a log-linear parameterization rather than a single power-law model. Unlike a power-law fit, which inherently enforces a strong monotonic decline even at small impact parameters, the log-linear form provides a more physically motivated description of the behavior observed in this regime.}

We perform a log-linear fit between $W_{2796}$ and $D$ for the GOTOQs, incorporating upper limits following the prescription of \citet{Rubin2018} and \citet{Dutta2020}. The likelihood function is defined as the product of the likelihoods for detections and non-detections,
\begin{equation*}
\begin{aligned} & \mathcal{L}(W) = \left(\prod^{n}_{i=1} \frac{1}{\sqrt{2\pi \sigma_i^2}} \rm{exp}\left\{ {-\frac{1}{2}\left[\frac{W_i - W(D_i)}{\sigma_i}\right]^2} \right\} \right) \\ & \times \left(\prod^{m}_{i=1} \int_{-\infty}^{W_i} \frac{dW^{\prime}}{\sqrt{2\pi \sigma_i^2}} \rm{exp} \left\{ -\frac{1}{2} \left[\frac{W^{\prime} - W(D_i)}{\sigma_i}\right]^2\right\} \right)
\end{aligned} \end{equation*}
where $W_i$ and $D_i$ are the $W_{2796}$ and impact parameter of the $i$-th measurement, respectively. The total error term is defined as $\sigma_i = \sqrt{\sigma_{int}^2 + \sigma_{mi}^2}$, accounting for both the intrinsic scatter, $\sigma_{int}$, and the measurement uncertainty, $\sigma_{mi}$. $W(D_i)$ denotes the model-predicted $W_{2796}$ at $D_i$.

The resulting best-fit relation is,
\begin{equation}
\log (W_{2796}/ \text{\AA}) = (-0.017 \pm 0.002) (D/ {\rm kpc}) + (0.547 \pm 0.019),
\end{equation}
with an intrinsic scatter of $\sigma_{int} = 0.965 \pm 0.037$ \AA. The best-fit log-linear relation is shown as a solid black line in Figure \ref{fig:width_vs_rho_all}, with the shaded gray region representing the $1\sigma$ uncertainty.

In summary, the GOTOQ sample densely populates the previously sparsely sampled regime at $D \lesssim 20$ kpc, providing an unprecedented view of the behaviour of  \MgII\ absorption at small galactocentric distances. The observed flattening of the canonical $W_{2796}$–$D$ anti-correlation in this regime indicates that the cool, metal-enriched gas traced by \MgII\ becomes increasingly saturated toward the inner halo. This smooth transition from the outer circumgalactic medium to extra-planar gas, and ultimately the interstellar medium suggests a continuous distribution of cold gas across these regions, rather than a sharp boundary. Our results therefore point to a coherent multiphase gas structure extending from the extended halo down to the disk-halo interface, with GOTOQs uniquely revealing this continuity at kiloparsec scales.

\subsection{Absorption-agnostic \gotoqs: \MgII\ Gas Covering Fractions at $D \lesssim$ 20 kpc}
\label{subsec:mgii_covering_fraction}

The \MgII\ covering fraction ($f_c$) is defined as the ratio of sightlines exhibiting \MgII\ absorption above a specified rest equivalent width (REW) threshold to the total number of sightlines. As expected, the covering fraction of cold gas traced by \MgII\ decreases rapidly with increasing impact parameter for isolated galaxies \citep{Dutta2020, Anand2021, Cherrey2025}, starting from $\sim20$ kpc. Directly probing the \MgII\ covering fraction at smaller separations ($D < 20$ kpc) has been challenging, since most quasar–galaxy pairs at such small angular separations have historically been identified through the very absorption features they are used to study. This limitation arises primarily because the bright background quasars often outshine their comparatively faint foreground galaxies, making independent identification difficult.

\begin{figure}
    \centering
    \includegraphics[width=0.99\linewidth]{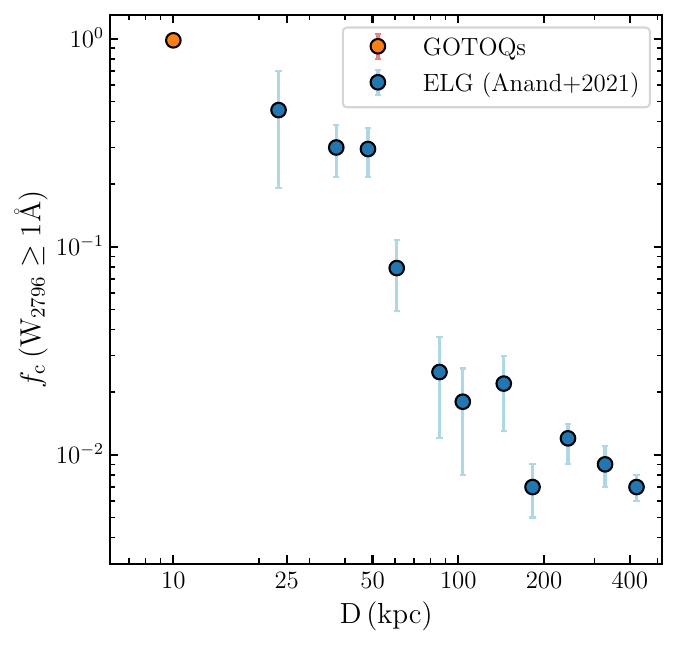}
    \caption{\MgII\ gas covering fraction around the emission line galaxies (ELG) as a function of the distances. Orange point corresponds to the \gotoqs, whereas the the blue points are taken from \citet{Anand2021}. }
    \label{fig:covering_frac}
\end{figure}

With our absorption-agnostic sample, we are now able to constrain, for the first time, the \MgII\ covering fraction below $D \lesssim 20$ kpc. Since the GOTOQs with photometric extensions have a median impact parameter of  11.6 kpc, and those without detectable extensions are likely to lie at even smaller separations, we treat our sample as probing the \MgII\ gas covering fraction over $0 < D \lesssim 20$ kpc, with an effective median separation of $\approx$10 kpc. To achieve this, we identified quasar sightlines in which the \MgII\ $\lambda2796$ transition lies within the spectral coverage and for which an \MgII\ absorption line with a rest equivalent width of 1\AA\ would be detectable at a significance of at least $3\sigma$, given the observed  signal-to-noise (SNR). This selection yielded 113 quasar sightlines. \MgII\ absorption was detected with $\rm W_{2796} \geqslant 1$\AA\ in all but two cases (SDSS~J220312.24+183446.4 with $z_{\rm gal} = 0.6583$ where no \MgII\ absorption ($3\sigma$ upper limit of $W_{2796} \leqslant 0.6$\AA) is detected, and SDSS~J104223.54+092708.2 with $z_{\rm gal} = 0.5922$ where $\rm W_{2796} = 0.59\pm0.15 $\AA), resulting in an estimated \MgII\ gas covering fraction of $f_c = 98.2^{+1.3}_{-4.5}$\% (95\% Wilson confidence interval) for the REW threshold of 1\AA\ and $D < 20$ kpc. In Figure \ref{fig:covering_frac}, we show the \MgII\ gas covering fraction with REW threshold of 1\AA\ around the emission line galaxies as a function of the projected physical distances. Orange point corresponds to the \gotoqs, whereas the the blue points are taken from \citet{Anand2021}, which studied \MgII\ gas covering fraction by cross-correlating $\approx 200,000$ emission line galaxies and the \MgII\ absorption systems present in the background quasar spectra from the SDSS-DR16 over the redshift range $0.5 < z < 1.0$.

\subsection{Line-of-sight Reddening}

It is well established that dust associated with foreground galaxies along the quasar line-of-sight can cause the quasar to appear redder. Quantifying this reddening is essential, as such dust-induced bias can influence quasar number counts and their selection in spectroscopic surveys. 

We estimate the line-of-sight reddening of background quasars due to the intervening galaxies in the \gotoqs\ sample assuming the Small Magellanic Cloud (SMC) extinction law \citep{Gordon_2003} and an average rest-frame spectral energy distribution (SED) of quasars given in \citet{Selsing2016}. The SMC extinction curve differs markedly from those of the Large Magellanic Cloud (LMC) and the Milky Way (MW), notably lacking the 2175~\AA\ absorption feature. Since this feature can be observed only for \gotoqs\ at redshifts $z \gtrsim 0.8$, and the majority of our sample lies below this threshold, we adopted the SMC-like extinction curve. Thus, the choice of extinction law has negligible impact on the derived results as LMC, MW and SMC extinction curves are similar in the rest wavelength range of interest here. We model the optical spectra of the background quasars using the composite quasar template redshifted to the quasar's rest frame. The SMC extinction curve was then applied at the redshift of the intervening galaxies, with the $V$-band extinction ($A_V$) as the only free parameter, along with a multiplicative scaling factor, following the procedure outlined by \citet{srianand2008a}. The resultant color excess $E(B-V)$ towards the GOTOQ sightlines ranges from $-0.261$ to $1.091$ with a median value of 0.087 mag (corresponding to median $A_V=0.238$, assuming $R_V=2.74$). 

\begin{figure}
    \centering
    \includegraphics[width=0.99\linewidth]{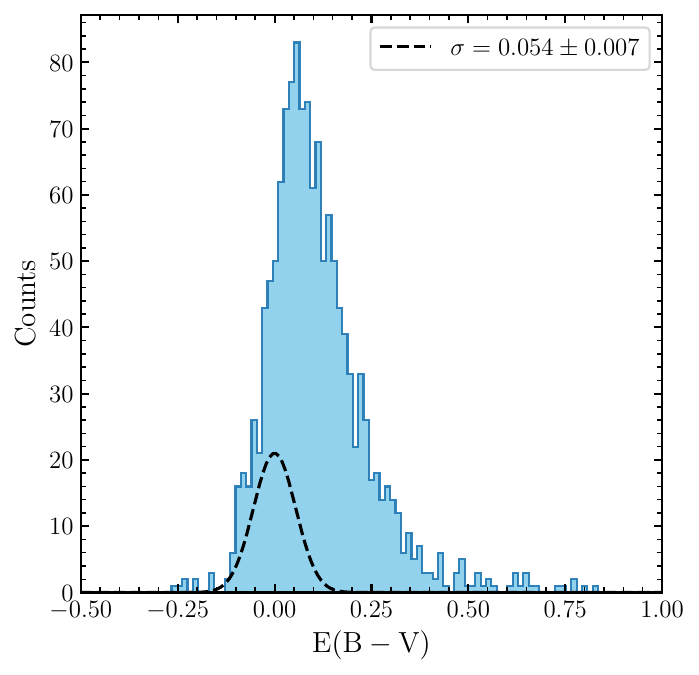}
    \caption{Histogram of the line-of-sight reddening of background quasars caused by the foreground galaxies, assuming an SMC-like extinction curve. The presence of negative E(B-V) values indicates possible systematic uncertainties of approximately 0.05 mag arising from quasar-to-quasar SED variations. This uncertainty is illustrated by the dashed black Gaussian curve centered at $E(B-V) = 0$, with a width of $\sigma = 0.054$.}
    \label{fig:reddening_all}
\end{figure}

{Histogram of the obtained color-excess towards the line-of-sight of all the \gotoqs\ in our sample is shown in Figure \ref{fig:reddening_all}}. The presence of negative E(B-V) values indicates possible systematic uncertainties of approximately 0.05 mag arising from quasar-to-quasar SED variations. This uncertainty is illustrated by the dashed black Gaussian curve centered at $E(B-V) = 0$, with a width of $\sigma \approx 0.05$ mag. Note that the median $E(B-V)$ derived from continuum fitting of the background quasars is substantially smaller than the value inferred from the Balmer decrement  (median $A_V\approx0.53$; see Section \ref{sec:ha_hb_ratio}). This indicates that, even at low impact parameters, extinction produced by the CGM dust is far less than the dust extinction associated with the star-forming regions traced by the Balmer lines (see section~\ref{sec:ha_hb_ratio}).

\begin{figure}
    \centering
    \includegraphics[width=0.99\linewidth]{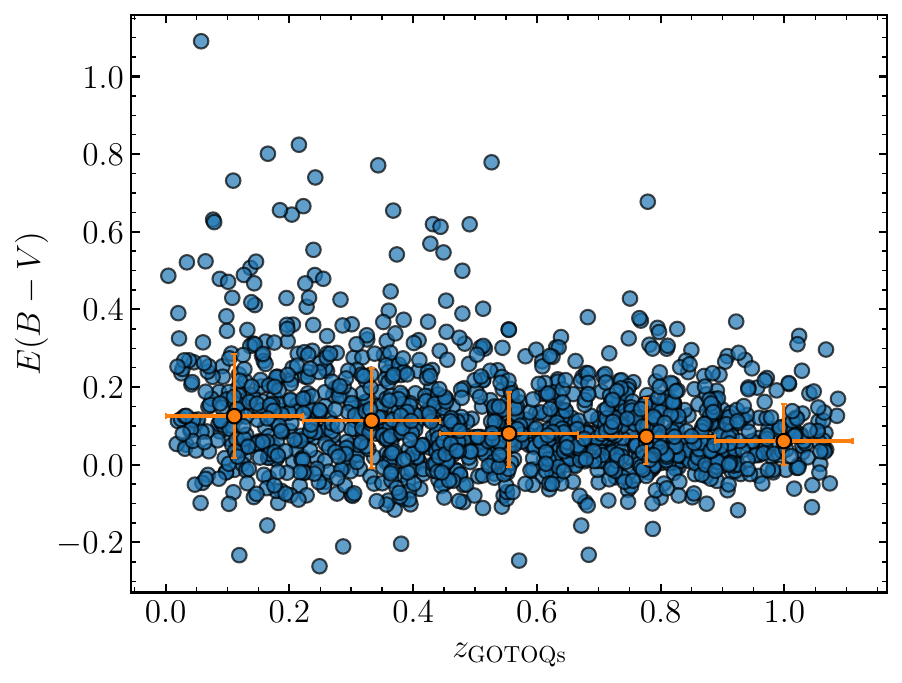}
    \caption{Redshift evolution of the line-of-sight reddening towards quasars caused by the foreground galaxies in the \gotoqs\ sample. The orange points indicate the median $E(B-V)$ values computed in redshift bins of width $\Delta z = 0.22$, while the error bars represent the 16th$-$84th percentile range within each bin.}
    \label{fig:ebv_evolution}
\end{figure}

Next, we compare the dust extinction due to the \gotoqs\ and that due to typical intervening galaxies probed by the \MgII\ and \CaII\ absorption systems as well as \HI\ 21 cm absorbers. Previous studies have established a strong correlation between color excess and absorber strength, with $E(B-V)$ increasing with $W_{2796}$ \citep{Menard2008}, empirically described by $E(B-V) = (8.0 \pm 3.0) \times 10^{-4}\, W_{2796}^{(3.48 \pm 0.30)}$ \citep{Budzynski2011}. For the median equivalent width of $W_{2796} = 2.15$\AA\ in our absorption-selected sample, the expected reddening is $E(B-V) \approx 0.012$ mag. In contrast, the \gotoqs\ sightlines exhibit substantially higher reddening than typical \MgII\ absorbers, likely because these quasars probe the dusty regions within or near the star-forming disks of their foreground galaxies. The \CaII\ absorbers, on the other hand, are known to comprise two distinct populations -- strong ($W_{3935} \geqslant 0.7$ \AA) and weak ($W_{3935} < 0.7$ \AA). The stronger systems are generally associated with smaller impact parameters and greater dust content \citep{wild2006}. The average reddening for strong \CaII\ absorbers, $E(B-V) \approx 0.099$, is comparable to that measured for the \gotoqs. Furthermore, \citet{Guha2025} recently reported a strong correlation between the integrated \HI\ 21 cm optical depth and line-of-sight reddening for \gotoqs, consistent with the trend observed in Galactic sightlines. Based on this relation and assuming a spin-temperature of 100 K, \gotoqs\ in our sample are expected to be damped \lya absorbers \citep[$N_{HI} \geqslant 2\times10^{20}\, \rm cm^{-2}$; see also][]{Kulkarni2022}, indicating that these systems are excellent probes of cold atomic and molecular gas and they can provide valuable insights into their cosmic evolution.

We next examine the redshift evolution of the line-of-sight reddening towards the background quasars in the \gotoqs\ sample. The blue points in Figure \ref{fig:ebv_evolution} show the color excess, $E(B-V)$, measured along each quasar sightline as a function of the redshift of the corresponding foreground galaxy. The orange points represent the median $E(B-V)$ values computed in redshift bins of width $\Delta z = 0.22$, with the vertical error bars indicating the 16th$-$84th percentile range within each bin. Although the individual measurements exhibit considerable scatter around the median trend, the binned data reveal a clear albeit a mild redshift dependence, with the average line-of-sight reddening decreasing at higher redshifts. A Spearman rank correlation test between $E(B-V)$ and $z_{\rm GOTOQs}$ yields a negative correlation ($r_S = -0.18$) with a near-zero $p$-value, confirming the statistical significance of this trend. This behavior is consistent with that observed for the intervening \MgII\ absorbers as the dust builds up in galaxies with time \citep{Napolitano2025}.

\subsection{Composite \gotoqs\ Spectrum: Detection of Diffuse Interstellar Bands}

Diffuse interstellar bands (DIBs) are generally weak optical or near-infrared absorption features that are widely attributed to large carbonaceous molecules or molecular ions. In Galactic sightlines, their strengths show a correlation, albeit with significant scatter, with the dust column density commonly expressed through the color excess, $E(B-V)$. Some of the strongest bands, particularly the broad $\lambda$4428 feature, have long been used as empirical tracers of diffuse, dust-rich gas \citep{Herbig1995, Kos2013}. The $\lambda$4428 band is both strong and broad, which makes it relatively easy to detect in high signal-to-noise spectra of reddened stars, although its width introduces uncertainties in the measurement of its centroid and profile.

Searches for DIBs beyond the Milky Way have revealed that several bands are present in nearby galaxies such as the LMC, SMC, M31, and M33 \citep{Ehrenfreund2002, Cordiner2008_M31, Cordiner2008_M33, vanLoon2013}. Their strengths are broadly consistent with Galactic relations once environmental factors and metallicity are taken into account \citep{Cordiner2008_M31, Cordiner2008_M33}. These extragalactic detections demonstrate that DIB carriers can survive outside the Milky Way, and that variations in ultraviolet radiation fields, metallicity, and grain processing influence the band strength per unit extinction. Extending DIB studies to more diffuse environments, such as the CGM using intervening quasar absorbers, has yielded only a few positive detections. A limited number of intervening systems show identifiable DIBs, including a $\lambda$4428 feature detected in the damped \lya absorber towards AO 0235+164 at $z\approx0.524$ \citep{Junkkarinen2004}, a $\lambda$6284 feature detected towards SDSS J163956.35+112758.7 at $z\approx 0.079$ and associated with a strong \HI\ 21-cm absorber \citep{srianand2013}, and the $\lambda$5780 band observed in a \CaII\ selected quasar sightline \citep{Ellison2008}. These detections typically require high signal-to-noise, targeted spectroscopy and are associated with sightlines characterized by high neutral hydrogen columns and significant dust content, conditions more typical of the interstellar medium than of diffuse halo gas.

\begin{figure*}
    \centering
    \begin{minipage}[b]{0.49\textwidth}
    \centering
    \includegraphics[width=0.99\linewidth]{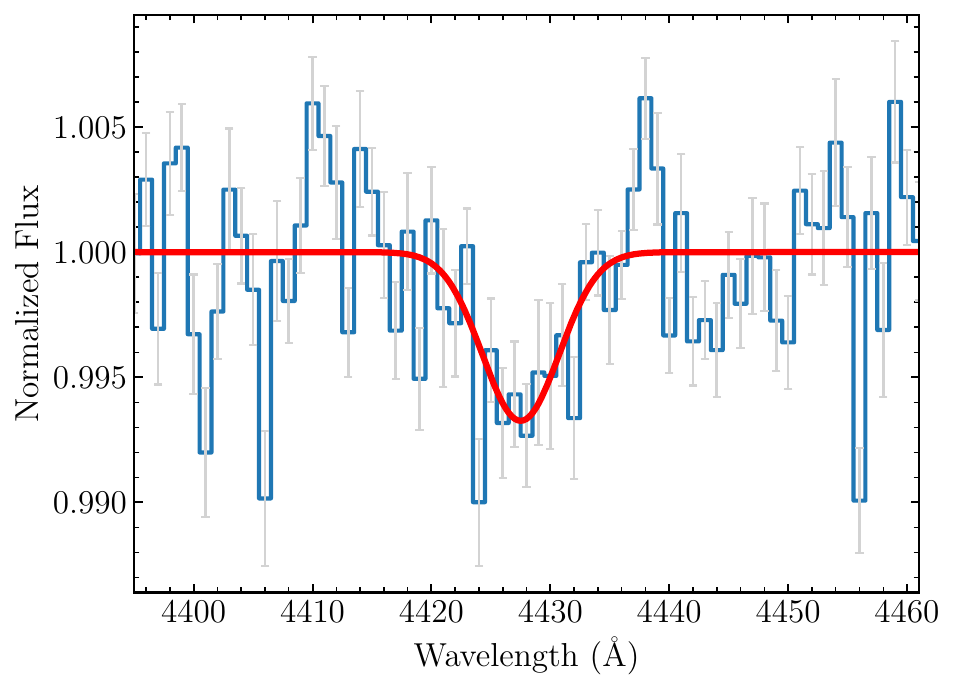}
    \end{minipage}
    \hfill
    \begin{minipage}[b]{0.49\textwidth}
    \centering
    \includegraphics[width=0.99\linewidth]{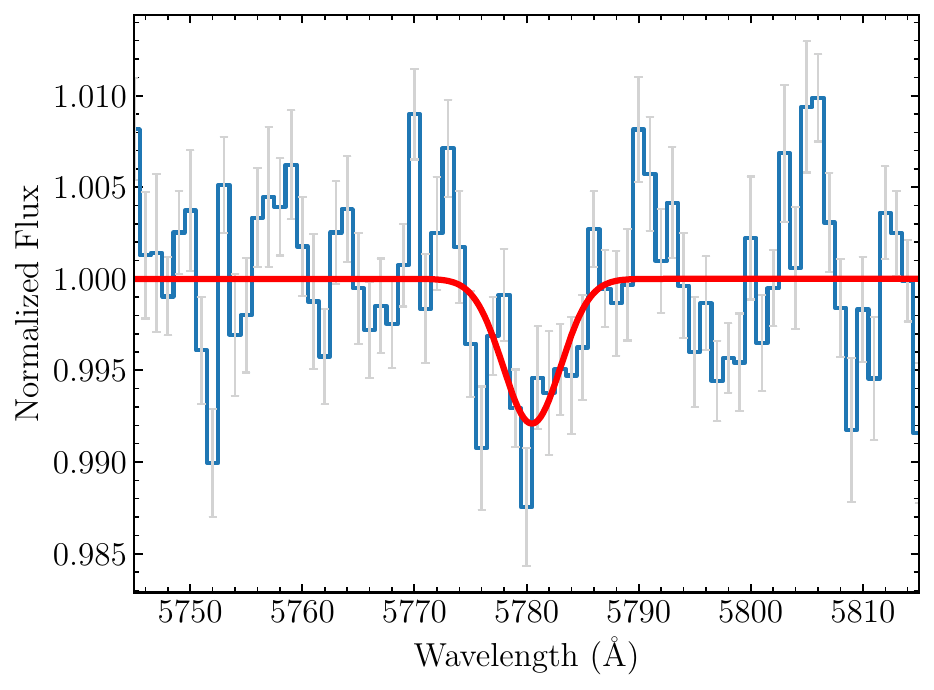}
    \end{minipage}
    \caption{ Detection of the DIB $\lambda4428$ (left panel) and DIB $\lambda5780$ (right panel) feature in the continuum-normalized, median-stacked spectrum (blue) of the \gotoqs\ in our sample. The solid red curve represents the Gaussian fit to the absorption feature observed in the stacked spectrum.
    }
    \label{fig:dib_4428}
\end{figure*}

More recently, large statistical searches using composite spectra of \MgII\ selected CGM sightlines have placed stringent limits on the presence of DIBs in halo environments. \citet{Chang2025} analyzed approximately 60,000 \MgII\ absorber spectra from SDSS to probe the DIB $\lambda$4428 at milli-angstrom precision and found no significant detection. Their composite spectra set upper limits at the few-milli-angstrom level, inconsistent with a simple extrapolation of the Galactic equivalent width$-$reddening relation into typical CGM dust columns. Overall, the literature indicates that DIB carriers are widespread in the interstellar medium of galaxies and detectable in several extragalactic ISM sightlines, but are largely absent or severely suppressed in \MgII\ traced CGM gas. 

We now investigate whether DIBs can be detected in the \gotoqs\ sample. Because DIB absorption features are intrinsically weak, spectra with very high signal-to-noise ratios (SNR) are required for a reliable detection. To improve the sensitivity, we construct a composite spectrum of the \gotoqs. Each quasar spectrum is first continuum-normalized and then shifted to the rest frame of the corresponding foreground galaxy. The spectra are subsequently resampled onto a common wavelength grid with a spacing of 1\AA, and the median flux is computed at each wavelength bin to obtain the composite \gotoqs\ spectrum. The uncertainty in the composite median spectrum is estimated using 200 bootstrap realizations, each generated by randomly replacing 20\% of the input quasar spectra. The standard deviation of the resulting flux values at each wavelength bin is adopted as the 1$\sigma$ error. The number of quasars contributing to the stack at any given wavelength varies depending on the absorber redshift, the quasar redshift (since regions blueward of the \lya\ emission line are excluded), and the observed spectral coverage of the spectra.

Figure~\ref{fig:dib_4428} presents the composite spectra in the regions around the expected DIB~$\lambda4428$ (left panel) and DIB~$\lambda5780$ features (right panel), with the corresponding $1\sigma$ uncertainties displayed in gray. {Around the DIB~$\lambda4428$ region, approximately 1336 quasar sightlines contribute to the stacked spectrum, whereas the DIB~$\lambda5780$ region is constructed from about 960 quasar sightlines owing to the more limited spectral coverage at longer rest-frame wavelengths}. In both cases, a broad depression in the composite flux is evident near the expected wavelength of the DIBs. The solid red curves show the best-fitting Gaussian profiles to these absorption features. The resulting rest-frame equivalent width of the DIB~$\lambda4428$ is $0.055 \pm 0.011$,\AA, corresponding to a detection significance of approximately $5\sigma$. We detect the DIB~$\lambda4428$ feature (albeit at a slightly lower significance) even after dividing the sample into two subsamples about the median $E(B-V)$ derived from the SED fitting. We also find that the absorption lines of ions such as Zn~{\sc ii}, Mn~{\sc ii}, Na~{\sc i}, and Ca~{\sc ii} are significantly stronger in the high-$E(B-V)$ subsample (e.g., the \CaII\ REW is $0.43\pm0.05$,\AA\ compared to $0.25\pm0.05$,\AA\ in the low-$E(B-V)$ subsample), consistent with the expected correlation between $E(B-V)$ and metallicity \citep{Dwek1998}. We detect the DIB~$\lambda5780$ feature at a significance of approximately $4\sigma$ (right panel of Figure~\ref{fig:dib_4428}), with a rest-frame equivalent width of $0.051 \pm 0.012$,\AA. The independent detection of this second prominent DIB provides additional support for the reality of the DIB~$\lambda4428$ detection. No statistically significant detections are found for any other DIB features in the stacked spectra.

In the Milky Way, the strength of the $\lambda4428$ DIB correlates with line-of-sight reddening, and \citet{Lan2015} quantify this relation as  $W_r (\text{\AA}) = (1.22 \pm 0.04)\, E(B-V)^{0.89 \pm 0.02}$. For the median reddening of our \gotoqs\ sample, $E(B-V)=0.087$, the expected equivalent width is $W_r \approx 0.139 \pm 0.008$~\AA, more than twice the value measured from the stacked GOTOQ spectrum. This suggests that either the disk-halo interface probed by the \gotoqs\ contains fewer DIB carriers than the Milky Way for a given $E(B-V)$ value, or the carriers are destroyed or ionized in the harsher halo conditions characterized by low densities and stronger radiation fields. Detailed individual spectroscopic detection of DIBs in a follow-up study of a sub-sample of GOTOQs may shed light on this issue.

\section{Discussion and Summary}
\label{sec:discussion}
With the identification of 1,345 unique Galaxies On Top Of Quasars (GOTOQs) from the SDSS-DR16 and DESI-EDR datasets, we have compiled the largest catalogue of quasar–galaxy pairs at projected separations of $D \lesssim20$~kpc to date, resulting in almost a fivefold increase over previous samples \citep{straka2013, Joshi2017}. {This unprecedented dataset provides a valuable resource for investigating the gas, dust, and star-forming properties of galaxies probed essentially through their own disks or inner halos.}

\subsection{Nature of GOTOQs and their host galaxies}

The emission-line properties of GOTOQs indicate that they trace predominantly normal, star-forming galaxies. The measured [O\,{\sc iii}]/[O\,{\sc ii}] ratios follow the same redshift evolution as field galaxies \citep[see Figure~\ref{fig:o3_o2_lineratios};][]{Khostovan2016}, while the median Balmer decrement of ${\rm H}\alpha/{\rm H}\beta = 3.33$ implies a dust extinction of $A_V \approx 0.5$ mag assuming the \citet{Calzetti2000} extinction curve (see Figure~\ref{fig:ha_hb_lineratios}). The absence of any significant redshift trend in the Balmer decrement up to $z \approx 0.5$ suggests that the dust attenuation properties of such galaxies remain broadly unchanged over this epoch. The [O\,{\sc ii}] luminosities, although affected by fiber aperture effects, span $7\times10^{36} - 2\times10^{42}$~erg~s$^{-1}$, consistent with sub-$L^\star_{[\mathrm{O\,II}]}$ galaxies at comparable redshifts \citep[][see Figure~\ref{fig:o2lum_hist}]{Comparat2016}. These findings confirm that GOTOQs preferentially select main-sequence star-forming systems rather than AGN-dominated or extreme emission-line galaxies. The systematic increase of $L_{[\mathrm{O\,II}]}$ with redshift primarily reflects the increasing physical area subtended by the spectroscopic fibers rather than intrinsic evolution, as demonstrated by the scaling with projected fiber area (see Figure~\ref{fig:z_vs_o2lum}).

\subsection{Gas content and covering fractions}

Our absorption-agnostic sample provides a direct measurement of the cold-gas covering fraction at the smallest galactocentric radii { of star forming galaxies}. For impact parameters $D < 20$~kpc, the Mg\,{\sc ii}~$\lambda2796$ covering fraction above a rest-equivalent-width threshold of 1\,\AA\ is $f_{\mathrm{c}} = 98.2^{+1.3}_{-4.5}$~per~cent, substantially higher than the values reported at larger separations \citep{Dutta2020, Anand2021}. This is consistent with the earlier findings that the occurance of \HI\ 21-cm absorption and DLAs in GOTOQs are very high \citep{Kulkarni2022, Guha2025}. All this confirm that the inner few tens of kiloparsecs around star-forming galaxies are almost completely filled with cool, metal-enriched gas. 

The GOTOQs preferentially host strong Mg\,{\sc ii} absorbers with median $W_{2796} = 2.15$~\AA, nearly twice that of the parent \MgII\ population (see Figure \ref{fig:width_dist}) . A positive correlation is observed between $L_{[\mathrm{O\,II}]}$ and $W_{2796}$ (see Figure \ref{fig:width_vs_o2lum}), consistent with previous stacking analyses \citep{Menard2011}. Although part of this trend may arise from reduced fiber losses at smaller impact parameters, it may as well be driven by an intrinsic link between star-formation-driven outflows and the enriched gaseous halos. Detailed investigation of the velocity distribution of absorbing gas with respect to the systemic redshift and down-the-barrel absorption towards the galaxies are needed to establish such a picture.

Using photometric identifications of host galaxies, we confirm that the canonical anti-correlation between $W_{2796}$ and impact parameter flattens for $D \lesssim 20$~kpc, implying that the \MgII\ transition becomes saturated in the dense disk–halo interface (see Figure \ref{fig:width_vs_rho_all}). Such a flattening has also been observed in smaller samples \citep{Kacprzak_2013, Guha2022b}.

\subsection{Dust reddening and the interstellar environment}

The median color excess towards the background quasars, $E(B-V)=0.087$, is significantly higher than the typical reddening in randomly selected \MgII\ absorbers \citep{Menard2008, Budzynski2011}. This confirms that GOTOQs probe the dusty, metal-rich interstellar regions of their host galaxies rather than diffuse halo gas. The derived reddening is comparable to that of strong \CaII\ absorbers \citep{wild2005}, supporting the notion that GOTOQs trace similar environments with high neutral-gas fractions. A statistically significant negative correlation between $E(B-V)$ and redshift suggests a gradual decline in dust content with cosmic time (see Figure \ref{fig:ebv_evolution}), consistent with trends seen in intervening absorbers \citep{Napolitano2025}.

The reddening measured using the SED fitting (i.e., $A_V\approx$ 0.23) is smaller than that obtained using the Balmer decrement (i.e., $A_V\approx 0.5$).  This is consistent with what has been seen in the case of star-forming galaxies \citep[for example,][]{Calzetti2000}, where nebular emission (and light from young massive stars) suffers more extinction compared to the light from old and relatively less massive stars. This is interpreted as young stars being still embedded in parent molecular clouds (for about the first 100 million years) and hence suffering more reddening. In our case, the quasar line-of-sight may be probing even less dustier regions compared to the typical ISM gas. Thus, the relatively smaller extinction seen towards these quasars may not be surprising.

\subsection{Diffuse Interstellar Bands in GOTOQs}

The composite spectrum of 1336~\gotoqs\ reveals a statistically significant detection of the broad DIB~$\lambda4428$ feature. As shown in the left panel of Figure~\ref{fig:dib_4428}, the stacked spectrum displays a wide, shallow depression at the expected wavelength, with a best-fitting profile yielding a rest-frame equivalent width of $0.055 \pm 0.011$~\AA\ ($5\sigma$ detection). This constitutes the first statistical detection of the $\lambda4428$ band in quasar-galaxy pairs at such small impact parameters, demonstrating that the carriers of DIBs survive in the inner halo and disk-halo interface of these galaxies. { This detection is supported by the presence of absorption at the expected wavelength of DIB $\lambda$5780 feature at $\sim$ 4$\sigma$ level.}

In the Milky Way, the strength of the $\lambda4428$ band scales with line-of-sight reddening. However, this relation yields an expected DIB absorption strength more than twice the value measured from the GOTOQ composite spectrum. This shortfall indicates that the disk-halo interface probed by GOTOQs contains fewer DIB carriers per unit dust than typical Milky Way diffuse sightlines. Possible explanations include destruction or ionization of the carriers in the lower-density, more strongly irradiated halo environment, reduced formation efficiency in lower-metallicity gas, or differences in dust composition relative to the Galactic ISM. Despite this suppression, the clear detection of the $\lambda4428$ band demonstrates that GOTOQs trace dust-rich, molecule-bearing gas and opens a new pathway to studying organic chemistry beyond the Milky Way. It also highlights the sensitivity of DIB carriers to environmental conditions in the inner CGM.

\subsection{Implications and future prospects}

The GOTOQs sample offers unique opportunities for follow-up studies of the multi-phase interstellar and circumgalactic media:
\begin{enumerate}
    \item \textbf{Cold-gas and molecular follow-up:} The high Mg\,{\sc ii} covering fraction and strong dust extinction imply excellent prospects for H\,{\sc i}~21-cm and molecular (e.g., OH, CO) absorption searches with upgraded Giant Metrewave Telescope (uGMRT), MeerKAT (Karoo Array Telescope), and future Square Kilometer Array (SKA) pathfinders \citep[see][]{Guha2025}.
    \item \textbf{Orientation and feedback diagnostics:} High-resolution imaging (with e.g., \textit{HST}, \textit{JWST}) can be used to relate absorber properties to disk inclination and azimuthal orientation, distinguishing inflow and outflow geometries \citep{kacprzak2012, Rubin2022}.
    \item \textbf{Dust and molecule evolution:} Combining optical extinction, DIB strengths, and \HI\ 21-cm optical depths will allow constraints on dust-to-gas ratios and the chemistry of complex molecules across cosmic time.
    \item \textbf{Emission–absorption synergy:} Spatially resolved IFU spectroscopy (with e.g., MUSE, KCWI) can recover total star-formation rates and metallicities of GOTOQ hosts, enabling joint studies of emission-line diagnostics and absorption-line kinematics.
\end{enumerate}

In summary, the present GOTOQ catalogue constitutes the largest and most complete census of quasar–galaxy pairs at $D \lesssim 20$~kpc. GOTOQs provide direct, dust-biased probes of the inner circumgalactic medium and the interstellar disks of galaxies. Their high Mg\,{\sc ii} covering fraction, strong dust reddening, and detection of diffuse interstellar bands demonstrate that they are exceptional laboratories for studying the cold, dusty, and molecular phases of gas at intermediate redshifts. The GOTOQs sample thus lays the foundation for future multi-wavelength investigations of gas flows, feedback, and dust evolution in galaxies.

\section*{Acknowledgement}
LKG acknowledges the use of the high-performance computing facility \textsc{pegasus} at IUCAA, Pune. This work made use of the following software packages: \textsc{NumPy} \citep{numpy2020}, \textsc{SciPy} \citep{scipy2020}, \textsc{Matplotlib} \citep{matplotlib2007}, \textsc{AstroPy} \citep{astropy:2018}, \textsc{dust\_extinction} \citep{Gordon2024}, \textsc{pymccorrelation} \citep{Privon2020}, and \textsc{spectres} \citep{Spectres}. {We thank the anonymous referee for the constructive comments and valuable suggestions, which have helped improve the manuscript.}

\section*{Data Availability}
The data used in this study are publicly available from the SDSS\footnote{\url{https://www.sdss.org/dr16/}} and DESI\footnote{\url{https://data.desi.lbl.gov/doc/releases/edr/}} surveys. { The catalog will be made available to interested readers upon reasonable request to the authors following publication of the paper.}

\bibliography{sample701}{}
\bibliographystyle{aasjournalv7}

\end{document}